\documentclass{svjour3}                     % onecolumn (standard format)
\usepackage{amssymb}
\usepackage{graphicx,epsfig}
\usepackage{natbib}
\bibpunct{(}{)}{;}{a}{}{,}
\makeatletter
\renewcommand*{\@biblabel}[1]{\hfill}
\makeatother

\journalname{SSRv}

\newcommand\beq{\begin{equation}}
\newcommand\eeq{\end{equation}}
\def\mug{\mu{\rm G}}

\begin{document}

\title{Nonthermal radiation mechanisms}

\author{Vah\'{e}~Petrosian \and
        Andrei~Bykov \and
        Yoel~Rephaeli}
  
\authorrunning{Petrosian et al.}
\titlerunning{Nonthermal radiation mechanisms}

\institute{Vah\'{e} Petrosian \at Department of Applied Physics, 
           Stanford University, Stanford, CA, 94305 \\
           Kavli Institute of Particle Astrophysics and Cosmology, 
	   Stanford University, Stanford, CA, 94305 \\
           \email{vahep@stanford.edu}
           \and 
	   Andrei M. Bykov \at 
	   A.F. Ioffe Institute for Physics and Technology, 194021
            St. Petersburg, Russia
           \and 
	   Yoel Rephaeli \at 
	   School of Physics and Astronomy, Tel Aviv University, 
	   Tel Aviv, 69978, Israel
	     }

\date{Received: 4 November 2007; Accepted: 11 December 2007}
	    
\maketitle

\begin{abstract}

In this paper we review the possible radiation mechanisms for the observed
non-thermal emission in clusters of galaxies,  with a primary focus on the radio
and hard X-ray emission. We show that the difficulty with the non-thermal,
non-relativisitic Bremsstrahlung model for the hard X-ray emission, first
pointed out by \citet{Petrosian2001} using a cold target approximation, is
somewhat alleviated when one treats the problem more exactly by including the
fact that the background plasma particle energies are on average a factor of 10
below the energy of the non-thermal particles. This increases the lifetime of
the non-thermal particles, and as a result decreases the extreme energy requirement,
but at most by a factor of three. We then review the synchrotron and so-called
inverse Compton emission by relativistic electrons, which when compared with
observations can constrain the value of the magnetic field and energy of
relativistic electrons. This model requires a low value of the magnetic field
which is far from the equipartition value. We briefly review the possibilities
of gamma-ray emission and prospects for {\sl GLAST} observations. We also present a
toy model of the non-thermal electron spectra that are produced by the
acceleration mechanisms discussed in an accompanying paper.

\end{abstract}

\section{Introduction}
\label{intro}

The observed signatures of the non-thermal (NT) activity in the intra-cluster
medium (ICM) were described in details  by \citealt{Rephaeli2008} - Chapter 5,
this volume and  \citealt{Ferrari2008} - Chapter 6, this volume.  Here we give a
brief summary. The first and least controversial radiative signature comes from
radio observations. Synchrotron emission by a population of relativistic
electrons is the only possible model for the production of this radiation. In
the case of the Coma cluster, the radio spectrum may be represented by a broken
power law \citep{Rephaeli1979}, or a power law with a rapid steepening
\citep{Thierbach2003}, or with an exponential cutoff \citep{Schlickeiser1987},
implying the presence of electrons with similar spectra. Unfortunately, from
radio observations alone one cannot determine the energy of the electrons or the
strength of the magnetic field. Additional observations or assumptions are
required. Equipartition or minimum total (relativistic particles plus fields)
energy arguments imply a population of relativistic electrons with Lorentz
factor $\gamma\sim10^4$ and magnetic field strength of $B\sim\mu$G, in rough
agreement with the Faraday rotation measurements \citep[e.g.][]{Kim1990}. 
\citet{Rephaeli1979} and \citet{Schlickeiser1987} also pointed out that the
electrons responsible for the radio emission, should also produce a spectrum of
hard X-ray (HXR) photons  (similar to that observed in the radio band),  via
inverse Compton (IC) scattering of the Cosmic Microwave Background (CMB)
photons. This emission  is estimated to be the dominant emission component
around 50 keV. Detection of HXR radiation could break the degeneracy and allow
determination of the magnetic field and the energy of the radiating electrons.
In fact, because the energy density of the CMB radiation (temperature $T_0$)
$u_{\rm CMB} = 4\times 10^{-13}(T_0/2.8\,{\rm K})^4$~erg\,cm$^{-3}$ is larger
than the magnetic energy density $u_B = 3\times 10^{-14}(B/\mu{\rm
G})^2$~erg\,cm$^{-3}$, one expects a higher flux of HXR than radio radiation.

As already described in the above mentioned papers by Rephaeli et al. and
Ferrari et al., recently there has been growing evidence for this and other
signatures of the NT activity. Excess HXR and extreme ultraviolet (EUV)
radiation are observed at the high and low ends of the usual soft X-ray (SXR)
thermal Bremsstrahlung (TB) radiation. Fig.~1 shows all the flux $\nu F (\nu)$
(or equivalently the energy density $\nu u(\nu) = 4\nu F (\nu)/c)$ of the above
mentioned and other radiation for the Coma cluster. However, for the excess
radiation not only the exact mechanisms are controversial but even their NT
nature is questioned. The observed spectra of the excess radiation often can be
fit by thermal spectra with  higher and lower temperatures than that needed for
the SXR observations with almost the same confidence as with a NT power law. The
most natural NT process for these excesses (specially for HXRs) is the IC
scattering of the CMB photons. However, the relatively high observed HXR fluxes
require a large number of relativistic electrons, and consequently a relatively
low magnetic field for a given observed radio flux. For Coma, this requires the
(volume averaged) magnetic field to be ${\bar B}\sim 0.1-0.3$~$\mu$G, while
equipartition gives ${\bar B}\sim 0.4$~$\mu$G and Faraday rotation measurements
give the (average line-of-sight) field of ${\bar B}\sim 3$~$\mu$G
\citep{Giovannini1993,Kim1990,Clarke2001,Clarke2003}. In general the Faraday
rotation measurements of most clusters give $B > 1$~$\mu$G; see e.g.
\citet{Govoni2003}. Because of this apparent difficulty, various authors (see,
e.g. \citealt{Ensslin1999,Blasi2000}) suggested that the HXR radiation is due to
non-thermal Bremsstrahlung by a second population of NT electrons with a power
law distribution in the 10 to 100 keV range. In what follows we examine the
merits and shortcomings of the mechanisms proposed to interpret these
observations. We first consider the EUV observations briefly and then address
the thermal and NT (IC and non-thermal Bremsstrahlung) radiation model for the
HXR observations.

\section{EUV emission}
\label{EUV}

The EUV excess in the 0.07 to 0.14 keV range was first detected by the Extreme
Ultraviolet Explorer from Coma \citep{Lieu1996} and some other clusters. There
are claimed detections of similar excess emissions in the 0.1 to 0.4 keV band by
{\sl Rosat}, {\sl BeppoSAX} and {\sl XMM-Newton}. The observational problems
related to the EUV and soft excesses are discussed by \citealt{Durret2008} -
Chapter 4, this volume. Initially, these excesses were attributed to thermal
emission by a cooler (${\rm k}T\sim 2$ keV) component, but there are several
theoretical arguments against this possibility, most notable is that the
expected line emission is not observed. The alternative model is the IC
scattering by CMB photons, which, in principle, can be easily fitted over the
small range of observations. However, this will require a population of lower
energy ($\gamma\sim 10^3$) electrons, indicating that the power law distribution
required for production of radio radiation must be extended to lower energies
with a power law index $p\sim 3$. This of course will mean an order of magnitude
more energy in electrons and it makes equipartition less likely (see also the
discussion at the end of Sect.~3.2.2)).

{\sl In summary, some of the observations of the EUV emission are widely
questioned and their theoretical modelling is quite problematic.}

\begin{figure}    
\begin{center}
\includegraphics[width=\hsize]{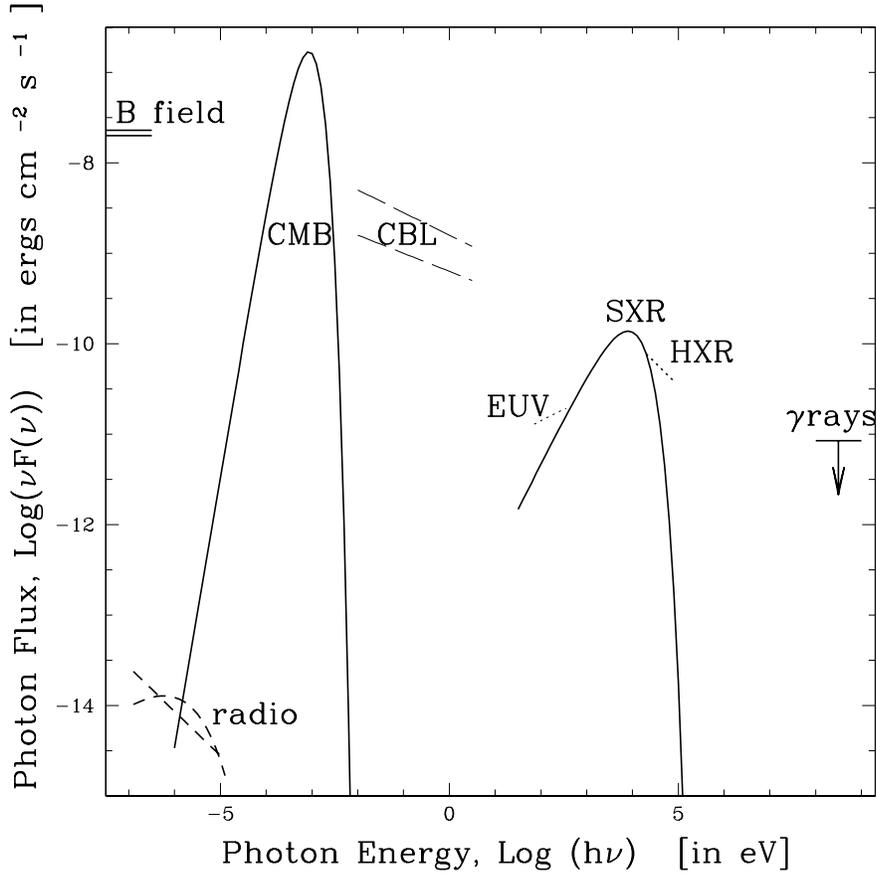}
\caption{The flux of all observed electromagnetic radiation for the Coma cluster
including cosmic microwave background (CMB), cosmic background light (CBL) and
static magnetic field (obtained from their energy density $u(\nu)$ as $\nu
F(\nu)= (c/4)\times \nu u(\nu)$). The spectra shown for the EUV and HXR range
are schematic and the upper limit in gamma ray range is from {\sl EGRET}
\citep{Sreekumar1996}. From \citet{Petrosian2003}.}
\label{photons}
\end{center}
\end{figure}

\section{Hard X-ray emission}

\begin{figure}
\begin{center}
\includegraphics[width=\hsize]{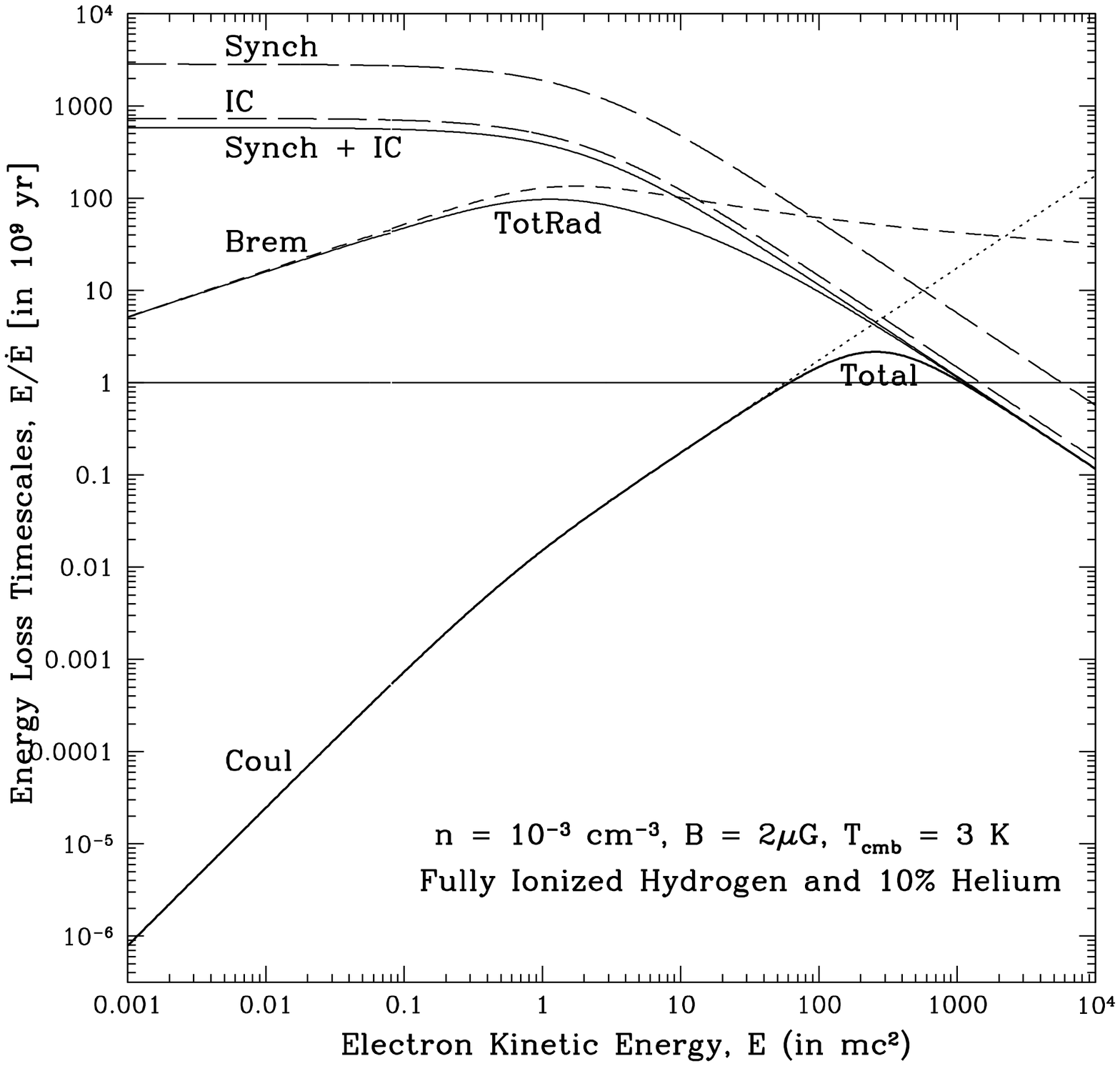}
\caption{Radiative, cold target Coulomb collision loss and other relevant
timescales as a function of energy for the specified  ICM conditions. The solid
lines from top to bottom give the IC plus synchrotron, total radiative and total
loss timescales. From \citet{Petrosian2001}).}
\label{timescales}
\end{center}
\end{figure}

The two possibilities here are Bremsstrahlung (thermal or non-thermal)
emission by a nonrelativistic electron population and IC emission by extreme
relativistic electrons which are also responsible for the observed radio
emission. There are difficulties in both cases but as we show below the IC is
the most likely interpretation. Fig.~\ref{timescales} gives the timescales for
the relevant processes for typical ICM conditions.

\subsection{Bremsstrahlung emission}

Bremsstrahlung radiation in the HXR (20 to 100 keV) range requires electrons of
comparable or somewhat larger energies. As pointed out by \citet[P01 for
short]{Petrosian2001}, and as evident from Fig.~\ref{timescales}, for such
nonrelativisitic electrons NT Bremsstrahlung losses are negligible compared to
elastic Coulomb losses. A thermal Bremsstrahlung interpretation of this emission
requires temperatures well above the virial value (see also below). Very
generally, for a particle of energy $E$ interacting with background electrons
and protons of much lower energy (cold target), the energy yield in
Bremsstrahlung photons $Y_{\rm brem}\equiv{\dot E}_{\rm brem}/{\dot E}_{\rm
Coul} = (4\alpha \ln\Lambda /3\pi)(E/m_{\rm e}{\rm c}^2)\sim 3\times
10^{-6}(E/25$~keV). Here $\alpha$ is the fine structure constant and
$\ln\Lambda$, the Coulomb logarithm, is $\sim 40$ for ICM conditions (for
details see \citealt{Petrosian1973}). Note that this result is independent of
the spectrum of emitting electrons which could be a Maxwellian of higher
temperature ($T_{\rm hot}\sim 30$~keV) or a power law of mean energy ${\bar
E}\gg kT$ of the background particles\protect\footnote{As shown below, also from the
acceleration point of view for this scenario, the thermal and NT cases cannot be
easily distinguished from each other. The acceleration mechanism energises the
plasma and modifies its distribution in a way that both heating and acceleration
take place.}. 
As pointed out in P01, for continuous production of a HXR luminosity of $L_{\rm
HXR}(\sim 4\times 10^{43}$ erg s$^{-1}$ estimated for Coma, see Table~1), a
power of $L_{\rm HXR}/Y_{\rm Brem}(\sim 10^{49}$ erg\,s$^{-1}$ for Coma) must be
continuously fed into the ICM. This will increase the ICM temperature to $T\sim
10^8$ K after $6\times 10^7$ yr, or to $10^{10}$ K in a Hubble time. An obvious
conclusion here is that the HXR Bremsstrahlung emission phase must be very short
lived.  It should also be noted that such a hot gas or such high energy
electrons cannot be confined in the ICM by gravity and will escape it in a
crossing time of $\sim 3\times 10^6$~yr unless it is confined by the magnetic
field or by scattering. From Fig.~\ref{timescales} we see that the Coulomb
scattering time, which is comparable to the Coulomb loss time, is equal to the
crossing time so that the escape time of the particles will be comparable to
these. Therefore, for confinement for periods exceeding the loss time we need a
shorter scattering mean free path or timescale. For example, for the scattering
timescale shown in this figure the escape time will be larger than the total
electron loss time at all energies.

These estimates of short timescales are based on energy losses of electrons in a
cold plasma which is an excellent approximation for electron energies $E\gg {\rm
k}T$ (or for $T_{\rm hot}\gg T$). As $E$ nears ${\rm k}T$ the Coulomb energy 
loss rate increases and the Bremsstrahlung yield decreases. For $E/{\rm k}T > 4$
we estimate that this increase will be at most about a factor of 3 (see below).

An exact treatment of the particle spectra in the energy regime close to the
thermal pool requires a solution of non-linear kinetic equations instead of the
quasi-linear approach justified for high energy particles where they can be
considered as test particles. Apart from Coulomb collisions, also collisionless
relaxation processes could play a role in the shaping of supra-thermal particle
spectra (i.e. in the energy regime just above the particle kinetic temperature).
We have illustrated the effect of fast collisionless relaxation of ions in a
post-shock flow in Figs. 1 and 2 of \citealt{Bykov2008b} - Chapter 8, this
volume.  The relaxation problem has some common features with N-body simulations
of cluster virialisation discussed in other papers of this volume where
collisionless relaxation effects are important. The appropriate kinetic
equations to describe both of the effects are non-linear integro-differential
equations and thus by now some simplified approximation schemes were used.

In a recent paper \citet{Dogiel2007} using linear Fokker-Planck equations for
test particles concluded that in spite of the short lifetime of the test
particles the ``particle distribution'' lifetime is longer and a power law tail
can be maintained by stochastic turbulent acceleration without requiring the
energy input estimated above. In what follows we address this problem not with
the test particles and cold plasma assumption, but by carrying out a calculation
of lifetimes of NT high energy tails in the ICM plasma. The proposed approach is
based on still linearised, but more realistic kinetic equations for Coulomb
relaxation of an initial distribution of NT particles. 

\subsubsection{Hot plasma loss times and thermalisation}

The energy loss rate or relaxation into a thermal distribution of high energy
electrons in a magnetised plasma can be treated by the {\sl{Fokker-Planck}}
transport equation for the gyro-phased average distribution along the length $s$
of the field lines $F(t,E,\mu,s)$, where $\mu$ stands for the pitch angle
cosine. Assuming an isotropic pitch angle distribution and a homogeneous source
(or more realistically integrating over the whole volume of the region) the
transport equation describing the pitch angle averaged spectrum, $N(E,t)\propto
\int\!\!\!\int F {\rm d}s{\rm d}\mu$, of the particles can be written as (see
\citealt{Petrosian2008} - Chapter 11, this volume for more details):
\beq\label{KEQCoul}
{\partial N \over \partial t} = {\partial^2\over \partial E^2} [D_{\rm Coul} (E) N] +
{\partial \over  \partial E }[{\dot E}_{\rm Coul} N],
\eeq
where $D_{\rm Coul}(E)$ and ${\dot E}_{\rm Coul}(E)$ describe the energy diffusion
coefficient and energy loss (+) or gain ($-$) rate due to Coulomb collisions. We
have ignored Bremsstrahlung and IC and synchrotron losses, which are 
negligible compared to Coulomb losses at low energies and for typical ICM
conditions.

As mentioned above, the previous analysis was based on an energy loss rate due
to Coulomb collisions with a "cold" ambient plasma (target electrons having zero
velocity):
\beq\label{cold}
{\dot E}^{\rm cold}_{\rm Coul} = m_{\rm e}{\rm c}^2/(\tau_{\rm Coul}\beta),\,\,\,\,\, 
{\rm where} \,\,\,\,\,
\tau_{\rm Coul}=(4\pi {\rm r}_0^2\ln\Lambda {\rm c}n)^{-1},
\eeq
$v = {\rm c}\beta$ is the particle velocity and ${\rm r}_0 = {\rm e}^2/(m_{\rm
e}{\rm c}^2 ) = 2.8\times 10^{-13}$~cm is the classical electron radius. For ICM
density of $n = 10^{-3}$~cm$^{-3}$, and a Coulomb logarithm $\ln\Lambda=40$,
$\tau_{\rm Coul} = 2.7\times 10^7$ yr. Note that the diffusion coefficient is
zero for a cold target so that ${\partial N\over \partial t} =\tau^{-1}_{\rm
Coul} {\partial\over \partial E} [N /\beta ]$ from which one can readily get the
results summarised above. As stated above this form of the loss rate is a good
approximation when the NT electron velocity $v \gg v_{\rm th}$, where $v_{\rm
th} = \sqrt{2{\rm k}T /m_{\rm e}}$ is the thermal velocity of the background
electrons. This approximation becomes worse as $v\rightarrow v_{\rm th}$ and
breaks down completely for $v < v_{\rm th}$, in which case the electron may gain
energy rather than lose energy as is always the case in the cold-target
scenario. A more general treatment of Coulomb loss is therefore desired.
\citet[PE07 herafter]{Petrosian2007} describe such a treatment. We summarise their
results below. For details the reader is referred to that paper and the
references cited therein.

Let us first consider the {\sl energy loss rate}. This is obtained from the rate
of exchange of energy between two electrons with energies $E$ and $E'$ which we
write as
\begin{equation}
\langle\Delta E\rangle /\Delta t = m_{\rm e}{\rm c}^2 G(E, E')/\tau_{\rm Coul}.
\label{exchange}
\end{equation}
Here $G(E,E')$ is an antisymmetric function of the two variables such that the
higher energy electrons loses and the lower one gains energy (see e.g.
\citealt{Nayakshin1998}). From their Eq.~24$-$26 we can write
\begin{equation} \label{cases}
G(E,E')=\cases{-\beta'^{-1},&if $E'>E, E\ll m_{\rm e}{\rm c}^2$;\cr
                \beta^{-1},&if $E'<E, E'\ll m_{\rm e}{\rm c}^2$;\cr
   m_{\rm e}{\rm c}^2(E'^{-1}-E^{-1}),&if $E, E'\gg m_{\rm e}{\rm c}^2$.}
\end{equation}
The general Coulomb loss term is obtained by integrating over the particle
distribution: 
\begin{equation}
\dot{E}_{\rm Coul}^{\rm gen}(E,t) ={m_{\rm e}{\rm c}^2\over \tau_{\rm
Coul}}\int\limits_0^\infty G(E,E')N(E',t){\rm d}E'.
\end{equation}
For non-relativistic particles this reduces to
\begin{equation}\label{gen}
\dot{E}_{\rm Coul}^{\rm gen}(E,t) ={\dot E}^{\rm cold}_{\rm Coul}\left(
\int\limits_0^E 
N(E',t){\rm d}E' - \int\limits_E^\infty (\beta/\beta')N(E',t){\rm d}E'\right).
\end{equation}
Similarly, the {\sl Coulomb diffusion coefficient} can be obtained from 
\begin{equation}
\frac{\langle(\Delta E)^2\rangle}{\Delta t}= 
\frac{(m_{\rm e}{\rm c}^2)^2 H(E, E')}{\tau_{\rm Coul}}
\label{exchange2}
\end{equation}
as $D_{\rm Coul}^{\rm gen}(E,t) = (m_{\rm e}^2{\rm c}^4 / \tau_{\rm Coul})
\int_0^\infty
H(E,E')N(E',t){\rm d}E'$.
From equations (35) and (36) of \citet{Nayakshin1998}\,
\footnote{Note that the first term in their equation (35) should have a minus sign and
that the whole quantity is too large by a factor of 2; see also
\citet{Blasi2000} for other typos.}
we get
\begin{equation} \label{casesdiff}
H(E,E')=\cases{\beta^2/(3\beta'),&if $E'>E, E\ll m_{\rm e}{\rm c}^2$;\cr
               \beta'^2/(3\beta),&if $E'<E, E'\ll m_{\rm e}{\rm c}^2$;\cr
		1/2,&if $E, E'\gg m_{\rm e}{\rm c}^2$.}
\end{equation}
Again, for nonrelativistic energies the Coulomb diffusion coefficient becomes 
\begin{equation}\label{gendiff}
D_{\rm Coul}^{\rm gen}(E,t) ={{\dot E}^{\rm cold}_{\rm Coul} m_{\rm e}{\rm c}^2 \beta^2\over
3}\left(\int\limits_0^E 
(\beta'/\beta)^2 N(E',t){\rm d}E' + \int\limits_E^\infty
(\beta/\beta')N(E',t){\rm d}E'\right).
\end{equation}

Thus, the determination of the distribution $N(E,t)$ involves solution of
the combined integro-differential equations Eq.~\ref{KEQCoul}, \ref{gen} and
\ref{gendiff}, which can be solved
iteratively. However, in many cases these equations can be simplified
considerably. For example, if  the bulk of the particles
have a Maxwellian distribution 
\begin{equation}
N(E') = n(2/\sqrt{\pi})({\rm k}T /m_{\rm e}{\rm c}^2
)^{-3/2}E'^{1/2}{\rm e}^{-E'/{\rm k}T},
\end{equation}
with ${\rm k}T\ll m_{\rm e}{\rm c}^2$, then integrating over this energy distribution, 
and after some algebra, the net energy loss (gain) rate and the diffusion
coefficient can be written as (see references in PE07):
\begin{equation} 
\dot{E}_{\rm Coul}^{\rm hot} = \dot{E}_{\rm Coul}^{\rm cold} 
     \left [{\rm erf}(\sqrt{x}) - 4 \sqrt{{x \over \pi}} e^{-x} \right ],
\label{hot} 
\end{equation}
and
\begin{equation}  
 D_{\rm Coul}(E) = \dot{E}_{\rm Coul}^{\rm cold} 
 \left({{\rm k}T \over m_{\rm e}
 {\rm c}^2}\right)
    \left[ {\rm erf}(\sqrt{x}) - 2 \sqrt{{x \over \pi}} e^{-x} \right],
\,\,\,\,\,\,{\rm with} \,\,\,\,\,\, x\equiv {E \over {\rm k}T},      
\label{DCoul} 
\end{equation}
where ${\rm erf}(x) = {2 \over \sqrt \pi} \int_0^x {\rm e}^{-t^2} {\rm d}t$ is the error
function.

\begin{figure}[hbtp]
\begin{center}
\includegraphics[width=\hsize]{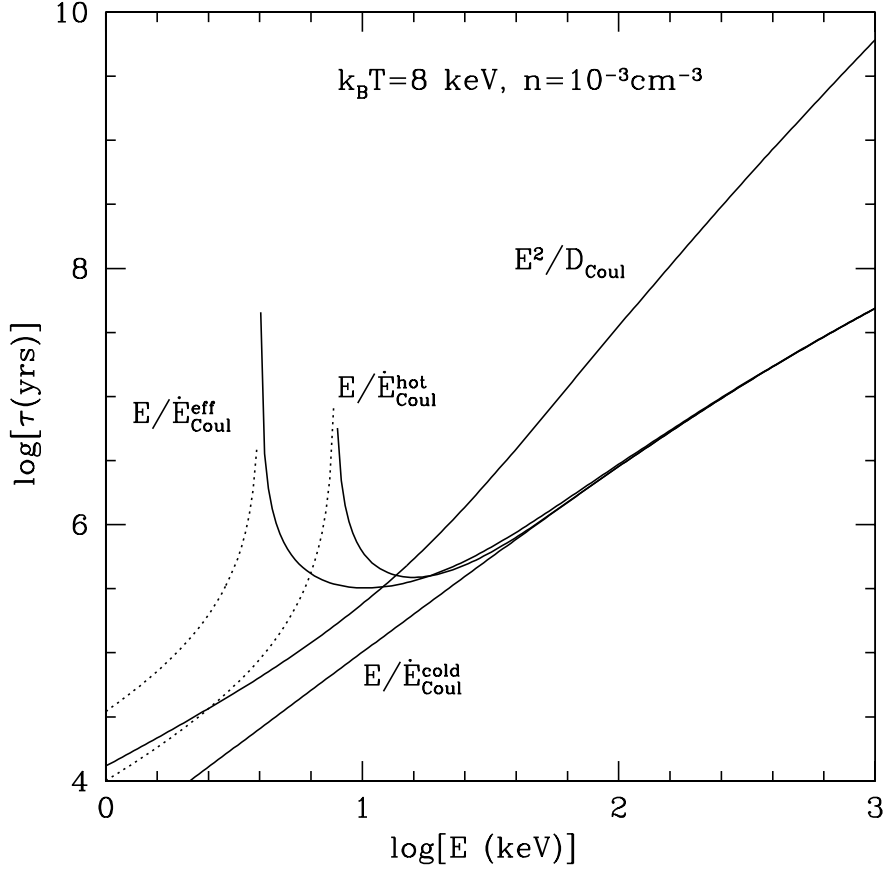}
\caption{
Various timescales for Coulomb collisions for hot and cold 
plasma with typical ICM parameters from Eqs.~\ref{cold}, \ref{hot}, 
\ref{DCoul} and \ref{eff}. Note that the
as we approach the energy $E\rightarrow {\rm k}T$ the loss time increases and
eventually becomes negative (or gain time) at low energies. The portions below
the spike show the absolute value of these timescales. From \protect\citet{Petrosian2007}.
}
\label{Coulcoeff}
\end{center}
\end{figure}

The numerical results presented below are based on another commonly used form of
the transport equation in the code developed by \citet{Park1995, Park1996},
where the diffusion term in Eq.~\ref{KEQCoul} is written as
${\partial\over \partial E}[D(E) {\partial\over \partial E}N(E)]$. This requires
modification of the loss term to
\begin{equation} \label{eff}
{\dot E}_{\rm Coul}^{\rm eff} = {\dot E}_{\rm Coul}^{\rm hot} + {{\rm d} D_{\rm Coul}
\over {\rm d} E} = \dot{E}_{\rm Coul}^{\rm hot} \left[ 1 - {1 \over x} {1 \over \gamma
(\gamma+1)} \right],
\end{equation}
where we have used Eqs.~\ref{hot} and \ref{DCoul}.
Fig.~\ref{Coulcoeff} shows the loss and diffusion
times, 
\begin{equation}
\tau_{\rm Coul} = E/{\dot E}_{\rm Coul}
\label{eqn:taucoul}
\end{equation}
and 
\begin{equation}
\tau_{\rm diff} = E^2/D_{\rm Coul}(E),
\label{eqn:taudiff}
\end{equation}
based on the above equations.

As a test of this algorithm PE07 show that  an initially narrow distribution of
particles (say, Gaussian with mean energy $E_0$ and width $\Delta E \ll E_0)$
subject only to  Coulomb collisions approaches a Maxwellian with ${\rm k}T =
2E_0/3$ with a constant total number of particles within several times the
theoretically expected thermalisation time, which is related to our $\tau_{\rm
Coul}$ as (see \citealt{Spitzer1962} or \citealt{Benz2002})
\beq\label{therm}
\tau_{\rm therm} = 3.5\tau_{\rm Coul}({\rm k}T /m_{\rm e}{\rm c}^2 )^{1.5} 
= 2\tau_{\rm Coul}(E_0/m_{\rm e}{\rm c}^2
)^{1.5}.
\eeq

\begin{figure}
\begin{center}
\includegraphics[width=0.48\textwidth]{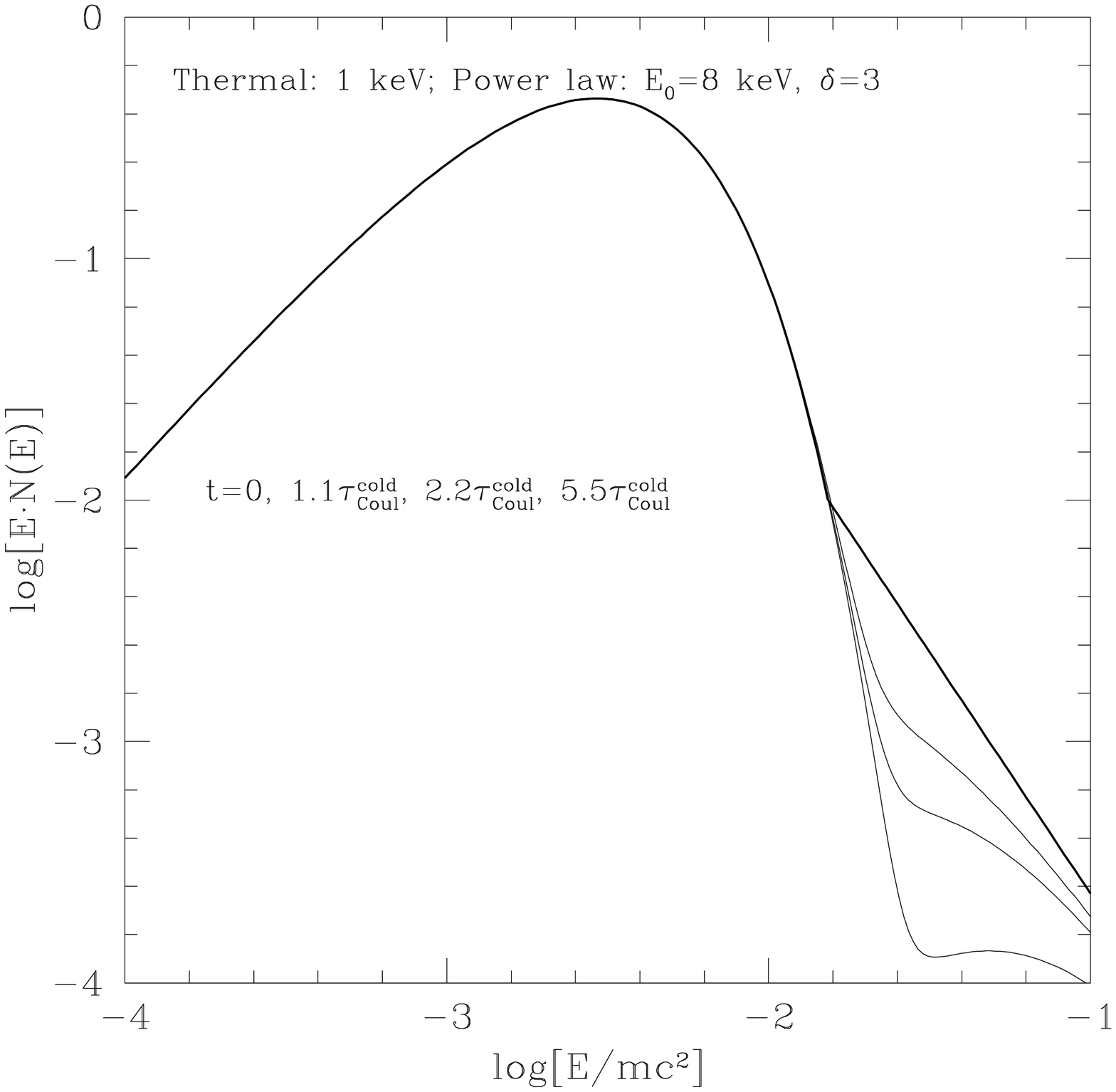}
\includegraphics[width=0.48\textwidth]{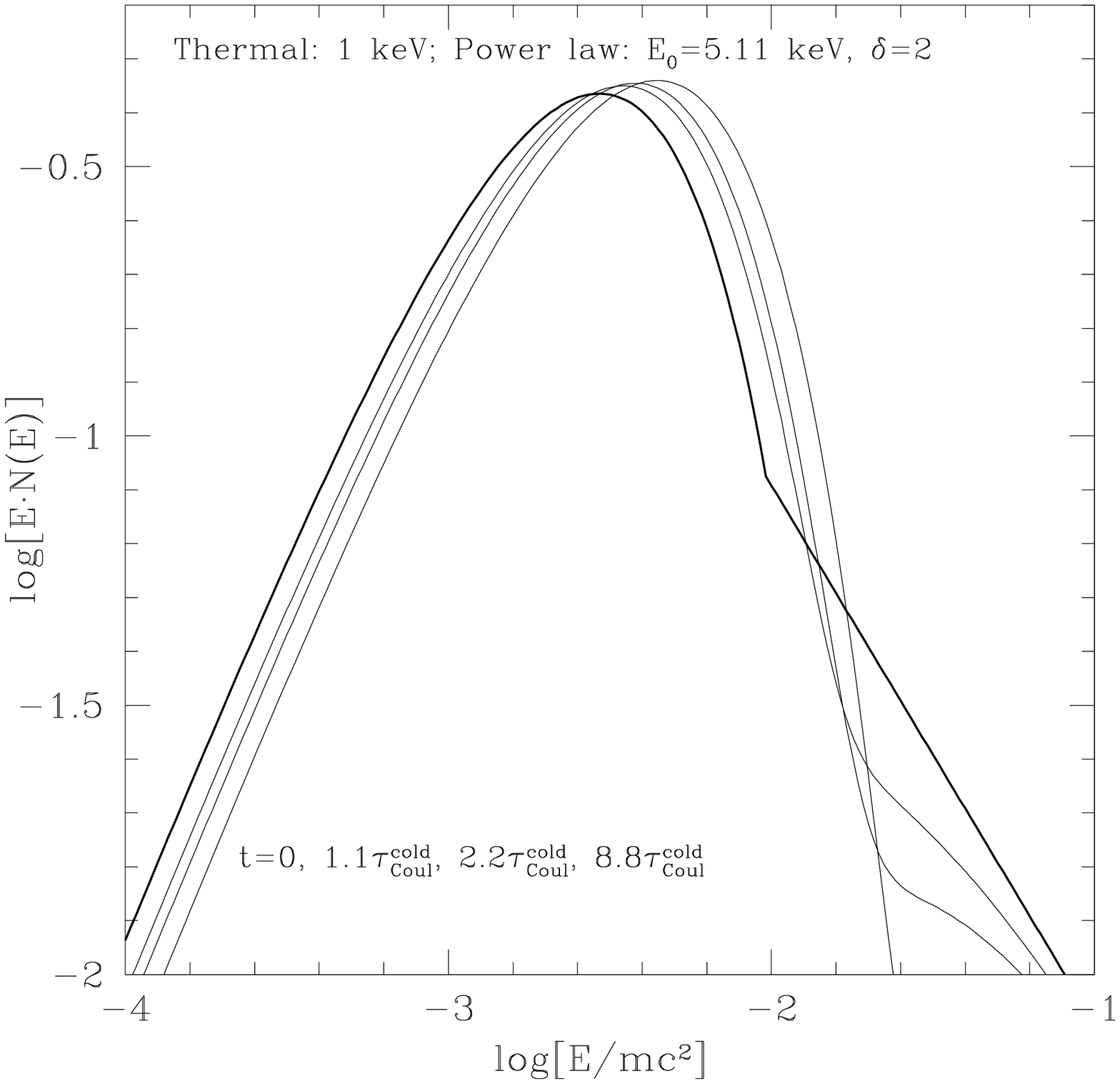}
\caption{Evolution of a NT power law tail (with isotropic angular distribution)
of electrons subject to elastic Coulomb collisions with background thermal
plasma electrons with initial temperature ${\rm k}T = 1$~keV, showing gradual
degradation of the power law NT tail (starting at energy $E_0$ with spectral
index $\delta$). {\sl Left panel:} Here we assume that the energy input by the
NT tail is negligible or carried away by some other means; i.e. the temperature
of the plasma is forced to remain constant. {\sl Right panel:} Here the energy
of the NT particles remains in the system so that  their thermalisation heats
the plasma. In both cases the NT tails are reduced by a factor of ten within
less than $100\times \tau_{\rm therm}$ or $<3$ times the cold target loss time
at the injected energy. From \cite{Petrosian2007}.
}
\label{relax}
\end{center}
\end{figure}

Using the above equations one can determine the thermalisation or the energy
loss timescale of supra-thermal tails into a background thermal distribution.
Fig.~\ref{relax} shows two examples of the evolution of an injected power-law
distribution of electrons. The left panel shows this evolution assuming a
constant temperature background plasma, which would be the case if either the
energy of the injected electrons was negligible compared to that of the
background electrons, or if the energy lost by the injected particles is lost by
some other means. The right panel does not make these assumptions and allows the
whole distribution to evolve. As is clearly seen in this figure, the energy lost
by the NT particles heats the plasma. It is also evident that the NT tail is
peeled away starting with low energies and progressing to higher ones. The NT
tail becomes negligible within less than 100 times the thermalisation time of
the background particles. These times are only about  three times larger than
the timescale one gets based on a cold target assumption for $E_0 = 20$ keV
particles. Similar conclusions were reached by  \citet{Wolfe2006}. The change in
the electron lifetime agrees also with Fig.~3 of \citet{Dogiel2007}. But our
results do not support the other claims in their paper about the longer lifetime
of the ``distribution of particles". The results presented here also seem to
disagree with \citet{Blasi2000}, where a procedure very similar to ours was used
(for more details see PE07).  

{\sl In summary the above results show that the conclusions based on the cold
plasma approximation are good order of magnitude estimates and that using the
more realistic hot plasma relations changes these estimates by factors of less
than three. The upshot of this is that the required input energy will be lower
by a similar factor and the time scale for heating will be longer by a similar
factor compared to the estimates made in P01.}

\subsection{Emission from relativistic electrons}

The radiative efficiency of relativistic electrons is much higher than that of
nonrelativisitic ones because of lower Coulomb losses  (see Fig.
\ref{timescales}). For ICM conditions electrons with GeV or higher energies lose
energy via synchrotron and IC mechanisms. At such energies the Bremsstrahlung
rate receives contributions from both electron-electron and electron-ion
interactions and its loss rate becomes larger than Coulomb rate but remains
below IC and synchrotron rates. Nevertheless, Bremsstrahlung emission  could
be the main source of gamma-ray radiation in the {\sl GLAST} range ($>$ GeV). In
this section we first compare the first two processes and explore whether the
same population of electrons can be responsible for both radio and HXR
emissions. Then we consider Bremsstrahlung and other processes for the gamma-ray
range. 

\subsubsection{Hard X-ray and radio emission}

As stated in the introduction, relativistic electrons of similar energies
($\gamma>10^3$) can be responsible for both the IC-HXR and synchrotron-radio
emission.  The IC and synchrotron fluxes depend on the photon (CMB in our case)
and magnetic field energy densities and their spectra depend on the spectrum of
the electrons (see e.g. \citealt{Rybicki1979}). For a power law distribution of
relativistic electrons, $N (\gamma) = N_{\rm total}(p -
1)\gamma^{-p}\gamma^{p-1}_{\min}$, and for $\gamma>\gamma_{\min}$, the spectrum
of both radiation components, from a source at redshift $z\equiv Z-1$ and
co-moving coordinate $r(Z)$, is given by
\beq\label{nufnu}
\nu F_i(\nu) = {\rm }c{\rm r}^2_0N_{\rm total}\gamma_{\min}
u_iA_i(p)(\nu/\nu_{{\rm cr},i})^{2-\alpha}
/(4\pi d^2_{\rm L}(Z),
\eeq
where $d_{\rm L}(Z)=({\rm c}/{\rm H}_0)Zr(Z)$ is the luminosity distance and $\alpha=(p+1)/2$ is
the photon number spectral index\,
\footnote{These expressions are valid for spectral index $p > 3$ or $\alpha >
2$. For smaller
indices an upper energy limit $\gamma_{\max}$ must also be specified and
the above expressions must be modified by other factors which are omitted here
for the sake of simplification.}. 
This also assumes that the relativistic
electrons and the magnetic field have similar spatial distribution so that the
spatial distribution of HXRs and radio emission are the same
(see e.g. \citealt{Rephaeli1979}). The clusters are unresolved at HXRs so at this time
this is the most reasonable and simplest assumption. A larger HXR flux will be
the result if the electron and $B$ field distributions are widely separated (see
e.g., \citealt{Brunetti2001}).

For synchrotron $\nu_{{\rm cr}, {\rm synch}} =
3\gamma_{\min}^2\nu_{B_\perp}/2$,  with $\nu_{B_\perp}={\rm e}B_\perp/(2\pi
m_{\rm e}{\rm c})$ and $u_{\rm synch}=B^2/(8\pi)$, and for IC, $u_i$ is the soft
photon energy density and  $\nu_{{\rm cr},{\rm IC}} = \gamma_{\min}^2\langle
h\nu \rangle$. For black body photons $u_{\rm IC} = (8\pi^5/15)({\rm
k}T)^4/({\rm hc})^3$ and $\langle {\rm h}\nu \rangle=2.8 \, {\rm k}T$.  $A_i$
are some simple functions of the electron index $p$ and are of the order of
unity and are given in \citet{Rybicki1979}.  Given the cluster redshift we know
the temperature of the CMB photons ($T=T_0Z$) and that $\nu_{{\rm cr},{\rm
IC}}\propto Z$ and $u_{\rm IC} \propto Z^4$ so that we have  
\beq\label{ratio}
{\cal R}=F_{\rm HXR}/F_{\rm radio}\propto Z^{2+\alpha }/B_\perp^{\alpha}.
\eeq
Consequently, from the observed ratio of fluxes we can determine the strength of
the magnetic field.

For Coma, this requires the {\sl{volume averaged}} magnetic field to be ${\bar
B}\sim 0.1 \, \mug$, while equipartition gives ${\bar B}\sim 0.4 \, \mug$ and
Faraday rotation measurements give the {\sl{average line of sight}} field of 
${\bar B}_{\rm l}\sim 3 \, \mug$ \citep{Giovannini1993, Kim1990, Clarke2001,
Clarke2003}. In general the Faraday rotation measurements of most clusters give
$B> 1\, \mug$; see e.g. \citealt{Govoni2003}. However, there are several factors
which may resolve this discrepancy. Firstly, the last value assumes a chaotic
magnetic field with a scale of a few kpc which is not a directly measured
quantity (see e.g., \citealt{Carilli2002})\, \footnote{The average line of sight
component of the magnetic field in a chaotic field of scale $\lambda_{\rm
chaos}$ will be roughly $\lambda_{\rm chaos}/R$ times the mean value of the
magnetic field, where $R$ is the size of the region.}. Secondly, the accuracy of
these results has been questioned (\citealt{Rudnick2003}; but see
\citealt{Govoni2004a} for an opposing point of view). Thirdly, as shown by
\citet{Goldshmidt1993}, and also pointed out by \citet{Brunetti2001}, a strong
gradient in the magnetic field can reconcile the difference between the volume
and line-of-sight averaged measurements. Finally, as pointed out by P01, this
discrepancy can be alleviated by a more realistic electron spectral distribution
(e.g. the spectrum with exponential cutoff suggested by \citet{Schlickeiser1987}
and/or a non-isotropic pitch angle distribution). In addition, for a population
of clusters observational selection effects come into play and may favour
Faraday rotation detection in high $B$ clusters which will have a weaker IC flux
relative to synchrotron. The above discussion indicates that the Faraday
rotation measurements are somewhat controversial and do not provide a solid
evidence against the IC model.

We now give some of the details relating the various observables in the IC
model. We assume some proportional relation (e.g. {\sl equipartition}) between
the energies of the magnetic field and non-thermal electrons

\beq\label{equip}
{\cal E}_{\rm e}=N_{\rm total}{p-1 \over p-2}\gamma_{\min} m_{\rm e}{\rm c}^2=\zeta {B^2\over 8\pi}{4\pi R^3\over
3},
\eeq
where $R=\theta d_{\rm A}/2$ is the radius of the (assumed) spherical cluster
with measured angular diameter $\theta$ and angular diameter distance $d_{\rm
A}(Z)=({\rm c}/{\rm H}_0)r(Z)/Z$, and equipartition with electrons is equivalent
to $\zeta = 1$.

From the three equations (\ref{nufnu}), (\ref{ratio}) and (\ref{equip}) we can
determine the three unknowns $B, {\cal E}_{\rm e}$ (or $N_{\rm total}$) and
$F_{\rm HXR}$ purely in terms of $\zeta, \gamma_{\min}$, and the observed
quantities (given in Table 1) $z, \theta$ and the radio flux  $ \nu F_{\rm
radio}(\nu)$. The result is\,
\footnote{Here we have set the Hubble constant ${\rm H}_0=70$
km\,s$^{-1}$\,Mpc$^{-1}$, the CMB temperature ${\rm T}_0=2.8$ K, and the radio
frequency $\nu = 1.4$ GHz.  In general $B^{2+\alpha} \propto {\rm
H}_0\nu^{\alpha-1}, N_{\rm total}\propto {\rm H}_0^{-3}$ and $F_{\rm HXR}\propto
{\rm H}_0^2{\rm T}_0^{2+\alpha}$. We have also assumed an  isotropic
distribution of the electron pitch angles and set  $B=B_\perp(4/\pi)$.}
\beq\label{B}
(B/\mug)^{\alpha+2}=0.20 \, \zeta^{-1} \left({F(1.4 \, {\rm GHz}) \over 
{\rm Jy}}\right)\left({5' \over \theta}\right)^3\left({10^4 \over
\gamma_{\min}}\right)^{2\alpha-3}
{Z^{3-\alpha}\over r(Z)},
\eeq
\beq\label{Ntot}
N_{\rm total} = 2.3\times 10^{65} \, {\alpha-3/2\over \alpha-1}\zeta\left({10^3 \over
\gamma_{\min}}\right)\left({\theta \over 5'}\right)^3
\left({B \over \mug}\right)^2\left({r(Z) \over Z}\right)^3,
\eeq
and
\beq\label{HXRflux}
\epsilon F_{\rm HXR}(\epsilon)=0.034\times F_0\left({N_{\rm total} \over
10^{65}}\right)
\left({10^4 \over \gamma_{\min}}\right)^{2\alpha-4}\left(
{\epsilon \over 5.9 \, {\rm keV}}\right)^{2-\alpha}\left({Z \over
r(Z)}\right)^2, 
\eeq
where we have defined $F_0\equiv 10^{-11}$ erg cm$^{-2}$ s$^{-1}$.
Note also that in all these expressions one may use  the radius of the cluster  
$R=3.39 \, {\rm Mpc} \, (\theta /5') \, r(Z)/Z$ instead of the angular radius
$\theta$.  

\begin{table}
\caption{Observed and estimated properties of clusters}
\vspace{0.3cm}
\begin{tabular}{|l|ccccc|cc|}
\hline \hline 
Cluster & $z$ & $kT\,^{a}$ & $F_{1.4 {\rm GHz}}\,^{b}$ &
$\theta\,^{c,b}$ & $F_{\rm SXR}$ 
& $B^{d}$ & $F_{\rm HXR}\,^{e}$ 
\\
 & & (keV) & (mJy)  & (arcmin) & ($F_0$)$\,^{f}$ & ($\mug$) &
($F_0$)$\,^{f}$
\\ 
\hline 
Coma   & 0.023 & 7.9 & 52 & 30 & 33 & 0.40 & 1.4 (1.6) 
\\ 
A 2256 & 0.058 & 7.5 &400 & 12 & 5.1 & 1.1 & 1.8 (1.0) 
\\
1ES 0657$-$55.8 & 0.296 & 15.6 & 78 & 5 & 3.9 & 1.2 & 0.52 (0.5) 
\\
A 2219       & 0.226 &12.4 & 81 & 8 & 2.4 & 0.86 & 1.0
\\
A 2163       & 0.208 & 13.8 & 55 & 6 & 3.3 & 0.97 & 0.50 (1.1) 
\\ 
MCS~J0717.5+3745& 0.550 & 13 & 220 & 3 & 3.5 & 2.6 & 0.76
\\
A 1914       & 0.171 & 10.7 & 50 & 4 & 1.8 & 1.3  & 0.22
\\
A 2744       & 0.308 & 11.0 & 38 & 5 & 0.76 & 1.0 & 0.41
\\ 
\hline
\end{tabular}
$^{a}${From \protect\citet{Allen1998}, except 1ES~$0657-55.8$ data from 
\protect\citet{Liang2000}} \\
$^{b}${Data for Coma from \protect\citet{Kim1990}; 
A~2256, A~2163, A~1914 and A~2744 from \protect\citet{Giovannini1999};
1ES~$0657-55.8$ from \protect\citet{Liang2000};
A~2219 from \protect\citet{Bacchi2003}.}\\
$^{c}${Approximate largest angular extent.}\\
$^{d}${Estimates based on equipartition. }\\
$^{e}${Estimates assuming $\zeta\gamma_{\min}=10^6$, with observed values
in parentheses: 
Coma \protect\citep{Rephaeli1999, Rephaeli2002, Fusco-Femiano1999, 
Fusco-Femiano2003}; 
Abell 2256 \protect\citep{Fusco-Femiano2000, Fusco-Femiano2004,Rephaeli2003}; 
1ES~$0657-55.8$ \protect\citep{Petrosian2006}; 
Abell~2163 \protect\citep{Rephaeli2006} these last authors 
also give a volume averaged $B\sim 0.4\pm 0.2 \mu$G.}\\
$^{f}${$F_0\equiv 10^{-11}$ erg cm$^{-2}$ s$^{-1} $}
\end{table}

\begin{figure}[htbp]
\begin{center}
\includegraphics[width=\hsize]{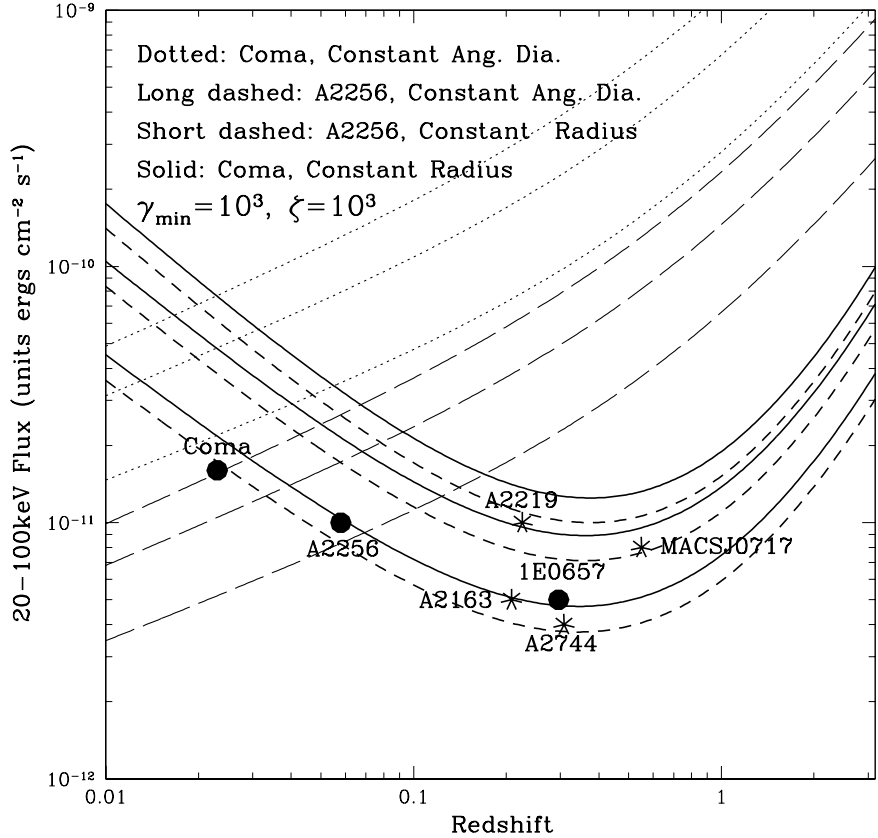}
\caption{Predicted variations of the HXR flux with redshift assuming a constant
physical diameter (solid and dashed lines) or constant angular diameter (dotted
and long-dashed lines), using the Coma cluster (dotted and solid) and A~2256
(long or short dashed) parameters assuming $\zeta=\gamma_{\min}=10^3$. In each
group, the IC photon spectral index $\alpha = 2.25, 2.0, 1.75$, from top to
bottom. Filled circles are based on observations and stars based on the
estimates given in Table 1. From \protect\citet{Petrosian2006}.)
}
\label{Fluxes}
\end{center}
\end{figure}

Table 1 presents a list of clusters with observed HXR emission and some other
promising candidates. To obtain numerical estimates of the above quantities in
addition to the observables $F_{\rm radio}, \theta$ and redshift $z$ we need the
values of $\zeta$ and $\gamma_{\min}$. Very little is known about these  two
parameters and how they may vary from cluster to cluster. From the radio
observations at the lowest frequency we can set an upper limit on
$\gamma_{\min}$; for Coma e.g., assuming $B\sim\mug$ we get
$\gamma_{\min}<4\times 10^3 (\mug/B)$. We also know that the cut off energy
cannot be too low because for $p>3$ most of the energy of electrons (${\cal
E}_{\rm e}=\int_{\gamma_{\min}} ^\infty\gamma^2N(\gamma)d\gamma$) resides in the
low energy end of the spectrum. As evident from Fig.~\ref{timescales} electrons
with $\gamma< 100$ will lose their energy primarily via Coulomb collisions and
heat the ICM. Thus, extending the spectra below this energy will cause excessive
heating. A conservative estimate will be $\gamma_{\min}\sim 10^3$. Even less is
known about $\zeta$. The estimated values of the magnetic fields $B$ for the
simple case of $\alpha=2$, equipartition (i.e. $\zeta=1$) and low energy cut off
$\gamma_{\min}=10^3$ are given in the 7th column of Table 1. As expected these
are of the order of a few $\mug$; for significantly stronger field, the
predicted HXR fluxes will be below what is detected (or even potentially
detectable). For $\alpha=2$ the magnetic field $B\propto
(\zeta\gamma_{\min})^{-1/4}$ and $F_{\rm HXR}\propto (\zeta\gamma_{\min})^{1/2}$
so that for sub-$\mu$G fields and $F_{\rm HXR}\sim F_0$ we need
$\zeta\gamma_{\min}\sim 10^6$. Assuming $\gamma_{\min}= 10^3$ and $\zeta=10^3$
we have calculated the expected fluxes integrated in the range of $20-100$ keV
(which for $\alpha=2$ is equal to $1.62 \times [20 \, {\rm keV} \, F_{\rm
HXR}(20 \, {\rm keV})$]), shown on the last column of Table 1. The variation of
this flux with redshift based on the observed parameters,  $\theta$ and $F_{\rm
radio}(\nu=1.4 \, {\rm GHz})$ of Coma and A~2256 are plotted in
Fig.~\ref{Fluxes} for three values of $\alpha=1.75, 2.0$ and 2.25 ($p=2.5, 3$
and 3.5) and assuming a constant physical radius $R$ which is a reasonable
assumption. We also plot the same assuming a constant angular diameter. This
could be the case due to observational selection bias if diffuse radio emission
is seen mainly from sources with $\theta$ near the resolution of the telescopes.
These are clearly uncertain procedures and can give only semi-quantitative
measures. However, the fact that the few observed values (given in parenthesis)
are close to the predicted values is encouraging. 

\begin{figure}[htbp]
\begin{center}
\includegraphics[width=\hsize]{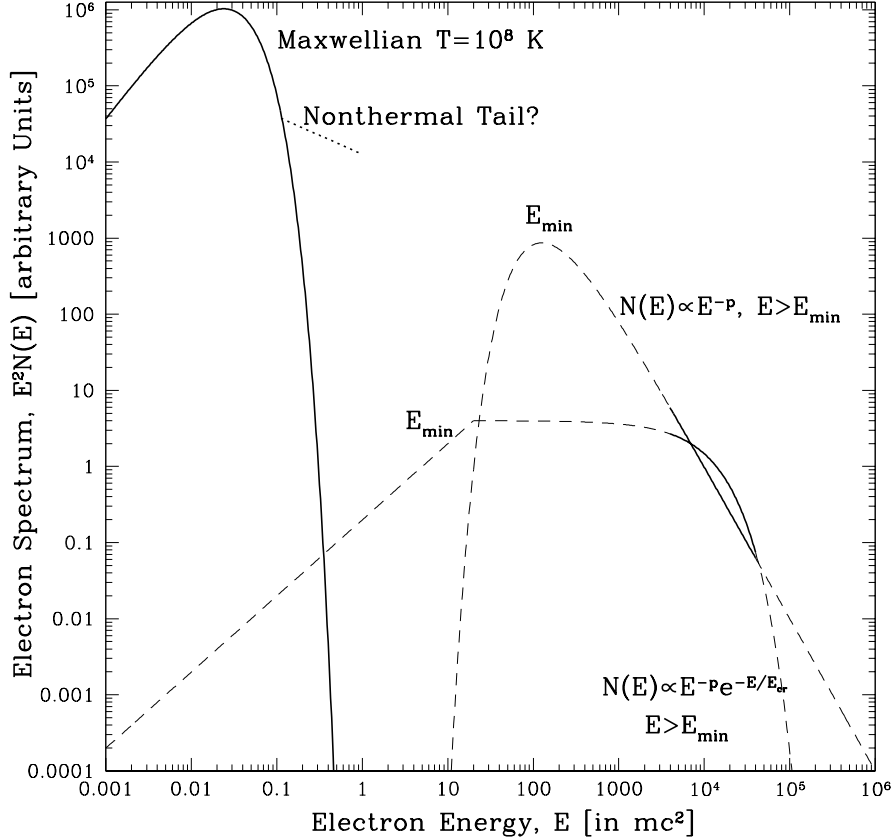}
\caption{The spectrum of electrons required for production of
radio and HXR
radiation based on the synchrotron and IC models. Values representing the Coma
cluster are used. The low energy peak shows the distribution of thermal
electrons along with a NT tail that would be required if the HXRs are
produced by the non-thermal Bremsstrahlung mechanism. The low end of the high energy NT electrons are
constrained to avoid excessive heating. The corresponding radio synchrotron
spectra are shown in Fig.~1. From \citealt{Petrosian2003}.
}
\label{electrons}
\end{center}
\end{figure}

{\sl In summary we have argued that the IC is the most likely process for
production of HXR (and possibly EUV) excesses  in clusters of galaxies. The high
observed HXR fluxes however imply that the situation is very far from
equipartition.} 

Our estimates indicate that $\zeta\sim 10^6/\gamma_{\min} > 10^3$. In addition
to the caveats enumerated above we note that, as described by
\citealt{Petrosian2008} - Chapter 11, this volume, the sources generating the
magnetic fields and the high energy electrons may not be identical so
that equipartition might not be what one would expect.

The spectrum of the required high energy electrons is best constrained by the
radio observations. In Fig.~1 we showed two possible synchrotron spectra. The
corresponding electron spectra are shown in Fig.~\ref{electrons} along with the
Maxwellian distribution of thermal electrons. The low end of the NT spectra are
constrained by requiring that their Coulomb loss rate be small so that there
will not be an excessive heating of the background particles. These span the
kind of NT electron spectrum that the acceleration model, discussed by
\citealt{Petrosian2008} - Chapter 11, this volume, must produce.

\subsubsection{Gamma-ray emission}

There have been several estimates of the expected gamma-ray flux, specifically
from Coma (see e.g. \citealt{Atoyan2000, Bykov2000, Reimer2004, Blasi2007}). The
{\sl EGRET} upper limit shown in Fig.~\ref{photons} does not provide stringent
constraints but {\sl GLAST} with its much higher sensitivity can shed light on
the processes described above. Several processes can produce gamma-ray
radiation. Among electronic processes one expects IC and Bremsstrahlung
radiation if electrons have sufficiently high energies.  If the electron
spectrum that accounts for radio and HXR emission can be extended to Lorentz
factors of $3\times 10^5$ (say with spectral index $p=3-4$) then the IC
scattering of CMB photons can produce 100 MeV radiation with a $\nu F(\nu)$ flux
comparable to the HXR fluxes ($\sim 10^{-11}-10^{-12}$ erg cm$^{-2}$ s$^{-1}$)
which can easily be detected by {\sl GLAST} (the former actually disagrees with
the {\sl EGRET} upper limit shown in Fig.~\ref{photons}). The lifetime of such
electrons is less than $10^7$ years so that whatever the mechanism of their
production (injection from galaxies or AGN, secondary electrons from
proton-proton interactions, see the discussion by \citealt{Petrosian2008} -
Chapter 11, this volume), these electrons must be produced throughout the
cluster within times shorter than their lifetime. A second source could be
electrons with $\gamma >10^4$ scattering against the far infrared background (or
cosmic background light) which will produce $>10$ MeV photons but with a flux
which is more than 100 times lower than the HXR flux (assuming a cosmic
background light to CMB energy density ratio of $<10^{-2}$) which would be
comparable to the {\sl GLAST} one year threshold of  $\sim 10^{-13}$ erg
cm$^{-2}$ s$^{-1}$. IC scattering of more abundant SXR photons by these
electrons could produce $>100$ MeV gamma-rays but the rate is suppressed by the
Klein-Nishina effect. Only lower energy electrons $\gamma< 10^2$ will not suffer
from this effect and could give rise to 100 MeV photons. As shown in
Fig.~\ref{electrons} this is the lowest energy that a $p=3$ spectrum can be
extended to without causing excessive heating (see P01). These also may be the
most abundant electrons because they have the longest lifetime (see Fig.
\ref{timescales}). The expected gamma-ray flux will be greater than the {\sl
GLAST} threshold.

High energy electrons can also produce NT Bremsstrahlung photons with energies
somewhat smaller than their energy. Thus, radio and HXR producing electrons will
produce greater than 1 GeV photons. Because for each process the radiative loss
rate roughly scales as ${\dot E}=E/\tau_{\rm loss}$, the ratio of the of
gamma-ray to HXR fluxes will be comparable to the inverse of the loss times of
Bremsstrahlung  $\tau_{\rm brem}$ and IC  $\tau_{\rm IC}$ shown in
Fig.~\ref{timescales} ($\sim 10^{-2}$ for $\gamma\sim 10^4$), which is just
above the {\sl GLAST} threshold.

Finally hadronic processes by cosmic rays (mainly $p-p$ scattering) can give
rise to $\sim 100$ MeV and greater gamma-rays from the decay of $\pi^0$
produced in these scatterings. The $\pi^\pm$ decays give the secondary electrons
which can contribute to the above radiation mechanisms. This is an attractive
scenario because for cosmic ray protons the loss time is long and given an
appropriate scattering agent, they can be confined in the ICM for a Hubble time
(see, \citealt{Berezinsky1997} and the discussion by \citet{Petrosian2008} -
Chapter 11, this volume). It is difficult to estimate the fluxes of this
radiation because at the present time there are no observational constraints on
the density of cosmic ray protons in the ICM. There is theoretical speculation
that their energy density may be comparable to that of the thermal gas in which
case the gamma-ray flux can be easily detected by {\sl GLAST}. \citet{Reimer2004}
estimate that {\sl GLAST} can detect  gamma-rays from the decay of $\pi^0$'s if
the cosmic ray proton energy is about 10~\% of the cluster thermal energy.

\section{Summary}

In this paper we address the processes that produce radio, EUV and HXR radiation
in the ICM. First we consider NT Bremsstrahlung as the source of HXRs which, as
shown earlier (P01), faces the difficulty of its low yield compared to Coulomb
losses. We describe results from a more detailed analysis of the lifetimes of NT
electron tails (or bumps) in a hot ICM than that presented in P01, where cold
target loss rates were used. We find that the lifetimes of NT tails is increased
by a factor of $<3$ so that the above difficulty becomes less severe but
production of HXRs via NT Bremsstrahlung remains problematical. Next we discuss
the expected radiative signature of relativistic electrons and show that radio
and HXR observations can be explained by synchrotron and IC scattering of CMB
photons. But now one requires a low magnetic field which is far from
equipartition with the electrons. Finally we give a rough estimate of the
gamma-ray signature of the relativistic electrons and point out several possible
scenarios in which the gamma-ray fluxes might exceed the {\sl GLAST} threshold.
Based on these results we present an average spectrum of electrons that is
required and possible extensions of it into the low energy regime. Production of
these spectra is discussed by \citealt{Petrosian2008} - Chapter 11, this volume.

\begin{acknowledgements}
The authors thank ISSI (Bern) for support of the team "Non-virialized X-ray
components in clusters of galaxies". A.M.B. acknowledges a support from RBRF 
grant 06-02-16844 and RAS Programs.
\end{acknowledgements}

\end{document}